# Tunable Band Alignment with Unperturbed Carrier Mobility of On-Surface Synthesized Organic Semiconducting Wires

*Andrea Basagni,[1] Guillaume Vasseur,[2,3] Carlo A. Pignedoli,[4] Manuel Vilas-Varela,[5] Diego Peña,[5] Louis Nicolas,[1,6] Lucia Vitali,[3,7] Jorge Lobo-Checa,[8,9] Dimas G. de Oteyza,[2,3,7,*] Francesco Sedona,[1,*] Maurizio Casarin,[1] J. Enrique Ortega,[2,3,10] Mauro Sambi.[1,11]*

[1] Dipartimento Scienze Chimiche Università Degli Studi Di Padova, Padova, Italy

[2] Donostia International Physics Center (DIPC), Paseo Manuel Lardizabal 4, E-20018 San Sebastián, Spain

[3] Centro de Física de Materiales (CSIC/UPV-EHU) -Materials Physics Center, Paseo Manuel Lardizabal 5, E-20018 San Sebastián, Spain

[4] NCCR MARVEL, Empa, Swiss Federal Laboratories for Materials Science and Technology, Dubendorf, Switzerland

[5] Centro de Investigación en Química Biolóxica e Materiais Moleculares (CIQUS) and Departamento de Química Orgánica, Universidade de Santiago de Compostela, 15782, Spain

[6] Ecole Normale Superieure de Cachan, Cachan, France

[7] Ikerbasque, Basque Foundation for Science, 48011 Bilbao, Spain

[8] Instituto de Ciencia de Materiales de Aragón (ICMA), CSIC-Universidad de Zaragoza, E-50009 Zaragoza, Spain

[9] Departamento de Física de la Materia Condensada, Universidad de Zaragoza, E-50009 Zaragoza, Spain

[10] Departamento de Fisica Aplicada I, Universidad del Pais Vasco, E-20018 San Sebastian, Spain

[11] Consorzio INSTM, Unità di Ricerca di Padova, Padova, Italy

ABSTRACT. The tunable properties of molecular materials place them among the favorites for a variety of future generation devices. Besides, to maintain the current trend of miniaturization of those devices, a departure from the present top-down production methods may soon be required and self-assembly appears among the most promising alternatives. On-surface synthesis unites the promises of molecular materials and of self-assembly, with the sturdiness of covalently bonded structures: an ideal scenario for future applications. Following this idea, we report the synthesis of functional extended nanowires by self-assembly. In particular, the products correspond to one-dimensional organic semiconductors. The uniaxial alignment provided by our substrate templates allows us to access with exquisite detail their electronic properties, including the full valence band dispersion, by combining local probes with spatial averaging techniques. We show how, by selectively doping the molecular precursors, the product's energy level alignment can be tuned without compromising the charge carrier's mobility.

Bottom-up covalent assembly is nowadays a versatile approach to synthesize low dimensional materials.[1-3] Its combination with well-defined substrates, typically termed as "on-surface chemistry", has achieved great progress over the past few years.[4,5]. Planar multifunctional molecular precursors with different size and shape have been used to grow networks with modular porosity and controlled connectivity,[6-9] moreover, the use of different functional groups and different substrates has afforded a significant increase of the degree of long range order through either stepwise or hierarchical approaches.[10-13] However, little attention and few experimental studies have been so far devoted to the study of the electronic properties of such surface supported polymers by averaging techniques.[14-16] To some extent this is because many of the produced surface-supported networks have a considerable level of structural defectiveness on the large scale and the probed species not always represent the majority product of the synthesis. Thus, intrinsically local probes such as scanning tunnelling microscopy (STM) and spectroscopy (STS) have been the primary tools for exploring these surface-confined products.[17-20] For this reason, although significant computational efforts have been devoted to investigate the electronic properties of 1D and 2D polymers,[21,22] experiments providing full band dispersion of such polymeric materials and thus a deeper understanding of their electronic properties are still scarce. Such properties of the polymer/substrate interface are, however, of outmost importance for the system's ultimate functionality. Their understanding and how to tune them by, for instance, chemical modifications, is thus a requisite towards the rational design and electronic structure engineering of functional materials and interfaces.

Poly-*para*-phenylene (PPP, see Figure 1) is a rigid-wire π-conjugated polymer that has attracted considerable attention. Its large band gap has been exploited to obtain blue light emitting diodes[23] and its conductivity can be easily increased through oxidative or reductive doping.[24] Besides, the booming interest in armchair graphene nanoribbons (A-GNR) has recently evidenced that PPP can be considered as the simplest A-GNR with N=3. [25,26] However, direct study of the electronic properties of pristine PPP, as well as its use in practical applications, have been hampered by its lack of processability. Oligomers longer than six phenyl rings are insoluble and infusible, and the material in both bulk and thin film forms is characterized by high defectiveness. [27] For these reasons until now the electronic properties of "infinite" PPP wires have been inferred from the properties of short, soluble (appropriately functionalized) or surface-supported oligomers,[19,28,29] rather than tested directly.

In this work we exploit on-surface synthesis to obtain extended and macroscopically ordered PPP wires and two different pyridinic derivatives. We report their full characterization by combined local (STM, STS) and surface averaging techniques [angle resolved photoemission (ARPES), core-level photoemission (XPS), near-edge X-ray absorption fine structure spectroscopy (NEXAFS) and low energy electron diffraction (LEED)], as well as by *ab initio* calculations based on density functional theory (DFT). Linking theory and experimental work, the energy-momentum dispersion of extended PPP molecular wires has been examined and their electronic structure dependence on nitrogen doping in specific sites of the organic scaffold has been probed.

**Results and discussion**

Figure 1 summarizes the adopted synthetic protocol and shows the chemical structure of the building blocks. Ullmann-like surface-assisted polymerization of brominated molecular precursors has been carried out to produce the polymers. Some of us have recently reported a detailed study on this synthesis protocol using 4,4"-dibromo-*p*-terphenyl (monomer **1** in figure 1) on Au(111) and the structural characterization of the final products: PPP wires and graphene nanoribbons.[26]

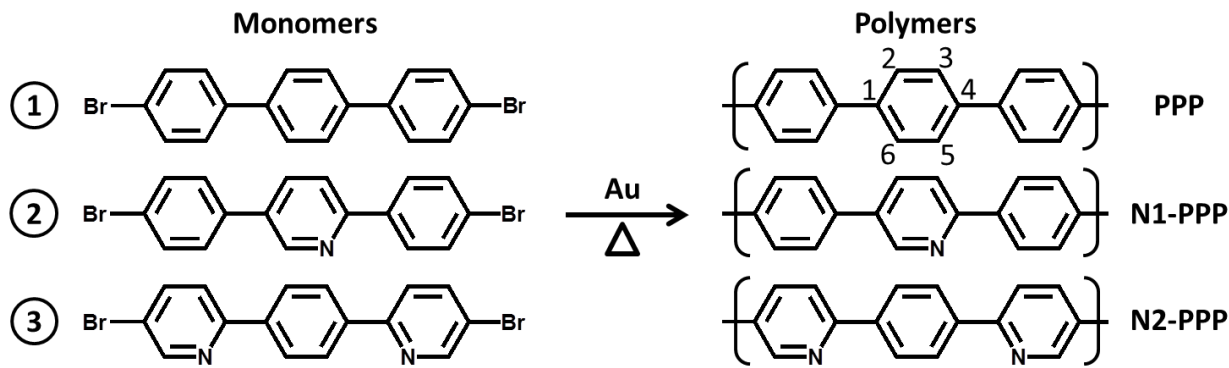

**Figure 1.** The molecular precursors with none (**1**), one (**2**) and two (**3**) pyridine rings are reported, along with the respective products PPP, N1-PPP and N2-PPP.

Since surface averaging momentum-resolved measurements require the macroscopic alignment of the PPP chains, the polymers were prepared by covalent coupling of 4,4"-dibromo-*p*-terphenyl (**1**) on clean Au(887). This surface is vicinal to the (111) plane and shows a periodic succession of narrow terraces (≈3.9 nm wide) with {111} plane orientation, whose monatomic steps run along the [1$\bar{1}$0] direction. Because PPP wires grow on Au(111) with their main axis parallel to the [1$\bar{1}$0] direction,[26] Au(887) allows one to align the polymers to obtain a highly anisotropic sample (see Figure 2a). As a result, the LEED pattern corresponds to a superstructure involving uniaxially aligned polymer chains commensurate with the substrate. A well-defined interchain spacing of ~10 Å along the [11$\bar{2}$] direction is evidenced by the LEED spots marked with blue arrows in Figure 2. This spacing supports the presence of Br atoms in between the polymer chains, also observed by STM. Along the chains we infer a periodicity of ~8.4 Å (0.75 Å$^{-1}$), twice the phenyl-phenyl distance, which coincides with thrice the Au lattice. An epitaxial matrix [4 2,0 3] is thus in excellent agreement with our LEED image. However, an unresolved striped motif is observed due to the lack of phase relation between the phenyl rings of adjacent chains (see Figure S1).

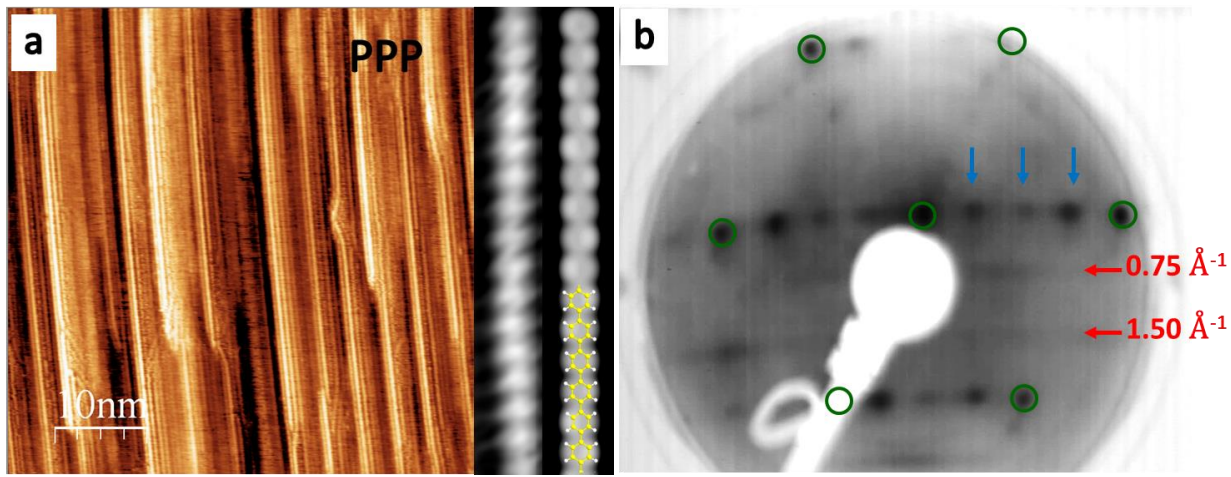

**Figure 2.** a) STM image (50x50 nm$^2$ V=0.60 V I=0.63 nA) of aligned PPP chains on Au(887) and comparison between experimental (V=-0.4V I=1.2nA) and DFT simulated STM images and b) experimental LEED pattern (E=50eV) of the same surface. Model color code: C yellow, H white.

The experimental band structures, E *versus* k, of the clean and PPP-covered Au(887) are illustrated in Figure 3a and 3b, respectively. From this direct comparison it is clear that the valence band with maximum (VBM) at E= -1.04 eV, $k_x$=1.49 Å$^{-1}$ and dispersing all the way to E=-6.3 eV at $\bar{\Gamma}$ originates from the polymeric wires (band marked with the green arrow in Figure 3b and in Figure S2, where the ARPES second derivative is displayed). The derived real space periodicity (4.21 Å) relates well with the expected inter-ring spacing of ≈4.3 Å (1.46 Å$^{-1}$)[30] assuming, as previously observed on *para*-sexyphenyl, that due to modulation of the photoemission transition matrices the measured band maximum corresponds to the polymer structure's second Brillouin zone. In addition, three non-dispersive bands at -2.7 eV, -4.6 eV and -6.22 eV are observed, marked by dotted lines in Figure 2b.

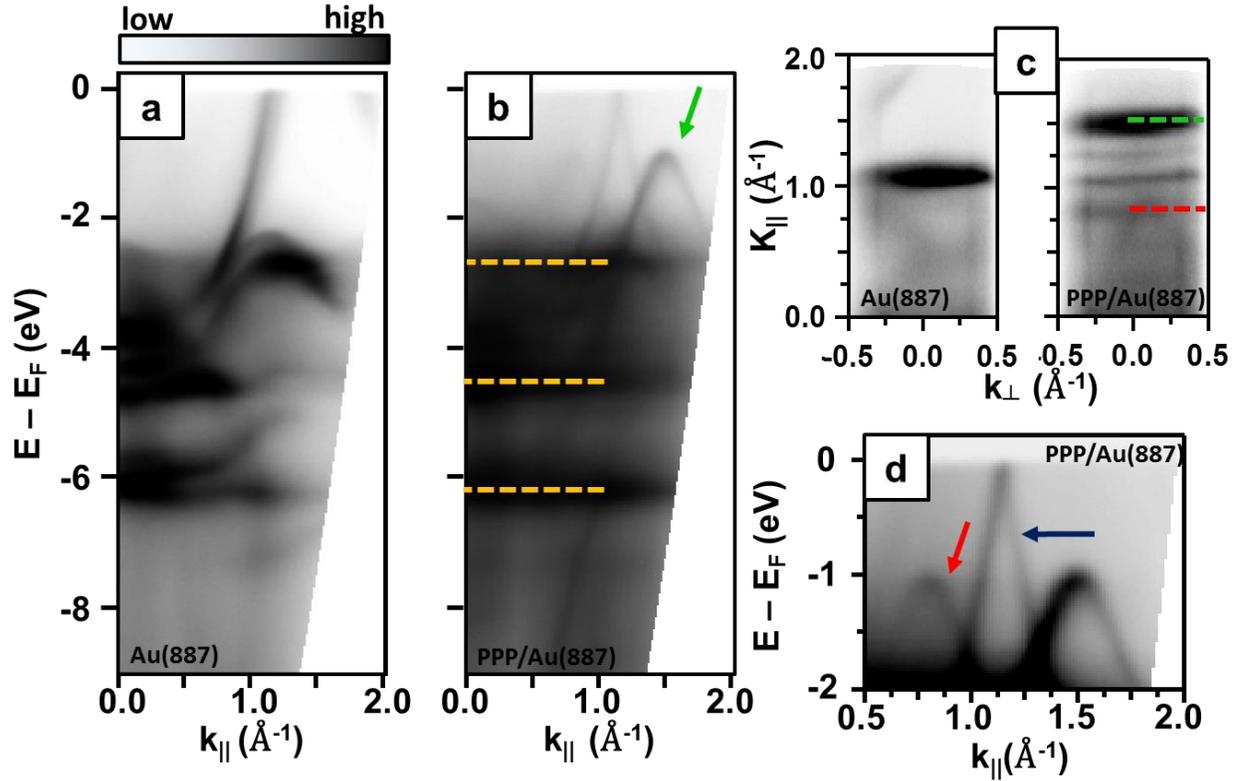

**Figure 3.** Angle resolved photoemission spectra (ARPES) I(E-E$_F$, k$_∥$), where k$_∥$ is along the [1$\bar{1}$0] direction (parallel to the wires), of a) clean Au(887) and of b) the oriented PPP/Au(887) interface. c) Constant energy map I(k$_∥$, k$_⊥$) at E-E$_F$=-1.04 eV of clean Au(887) (left) and of the oriented PPP/Au(887) (right) d) zoom-in of the spectra in b). Arrows and dashed lines highlight peculiar features, see text for details.

DFT calculated band structure (reported in Figure S3) performed on free-standing polymers, as well as on Au(111)-supported PPP wires with the experimentally found geometry, reproduces correctly the main features of experimental ARPES, such as the curvature of the main dispersive band and the presence of quasi flat bands at energies comparable with the experiment. Calculations with and without the Br atoms remaining in between chains show that the linear arrangement of Br causes none of the bands mentioned above. Figure SI3 also evidences that the dispersive band can be rationalized by a linear combination of benzene orbitals with the major weight at the linking carbon atoms (*i.e.* atoms 1 and 4 of each ring, related to benzene's HOMO) resulting in a molecular orbital delocalized along the PPP wire length, whereas the quasi-flat bands stem from localized molecular orbitals resulting from linear combination of ring-localized benzene orbitals (*i.e.* atoms 2, 3, 5 and 6 of each ring, related to benzene's HOMO-1). Moreover it is important to remark that the energies of the three experimentally observed non-dispersive bands coincide with the intensity maxima of the clean substrate 3*d* bands (see Figure 3a and 3b). Therefore the origin of these experimentally observed bands may be related to a combination of both substrate and PPP overlayer contributions.

Differently from the case of sexyphenyl, where a discretized dispersion has been observed,[29] our on-surface polymer shows a continuous dispersion of the band due to the overlap of a higher number of monomer orbitals. A theoretical study of the electronic structure dependence on the length in polyphenylene oligomers has shown that for a molecule with 18 phenyl rings the energy difference between successive states is so small that the situation is equivalent to the formation of a continuous band.[31] Hence our samples (Fig. 2a, Fig. S4), where the typical chain length greatly exceeds that length (18 phenyl rings correspond to only about 8 nm), can be regarded as a good representation of the electronic structure of an infinitely long PPP chain.

Constant energy maps of the clean surface and of PPP wires at the VBM (Figure 3c) reveal that its intensity is spread almost uniformly in $k_\perp$. That is, perpendicular to the wires (along [11-2]) the bands show negligible energy dispersion. Such flat bands imply that these orbitals are localized within the wires, due to a negligible intermolecular coupling. The result therefore reflects the occurrence of a truly one-dimensional band dispersion excluding electronic coupling between PPP wires.

Additional weaker features are observed at lower momentum values, see the red arrow in Figure 3d. In particular, a replica of the main dispersive band of the polymer lies at $k_{\parallel} \approx 0.75$ Å$^{-1}$, *i.e.* at half of its Brillouin zone ($\pi/4.2$ Å). This is in excellent agreement with the system's periodicity as measured with LEED and may be understood as stemming naturally from the epitaxial relationship with the substrate (which shows commensuration at every second phenyl ring). However, the fact that the unit cell in the direct space is twice the Ph-Ph separation may also be interpreted by a twisted conformation of the PPP chain due to the steric hindrance between the *orto* hydrogen atoms of neighboring phenyl rings.[32,33] Since the interaction with the substrate usually tends to flatten the molecules (in fact PPP chains are flat on Cu(110)[34] while sexyphenyl is twisted in thick films[29]), the twisted conformation should be confirmed by independent means.

Polarization-dependent NEXAFS measurements have been performed to gather insight into the adsorption geometry of the PPP chains. The spectra were acquired by scanning the [11$\bar{2}$] surface direction (perpendicular to chains) with the X-ray polarization projection. As observed in Figure 4a, the C K-edge exhibits four peaks at 285.07 ($\pi_1^*$), 288.97 ($\pi_2^*$), 293.69 ($\sigma_1^*$) and a broad peak at 301.59 eV ($\sigma_2^*$), in agreement with NEXAFS spectra collected from benzene physisorbed on metal surfaces.[35,36] The tilt angle of the $\pi^*$ orbital of the benzene ring relative to the surface normal has been obtained by comparing the angular dependent intensity of the $\pi_1^*$ resonance and the predicted resonance intensity according to a Stöhr-derived equation for twofold substrate symmetry,[37] see SI for details. As reported in Figure 4, the $\pi_1^*$ resonance intensity shows an opposite polarization dependence to that of $\sigma_1^*$ and $\sigma_2^*$ and it is maximized at grazing incidence, while the non-vanishing intensity at normal incidence implies a non-planar geometry: the derived orientation of the benzene ring planes relative to the surface is α=20°±5° (see Figure SI4). Therefore, NEXAFS linear dichroism confirms that the PPP chains adopt a twisted conformation wherein adjacent phenyl rings are alternately rotated clockwise and counterclockwise with respect to the PPP main axis. The full twist angle of 40°±10° is of the order of that predicted for PPP in the gas phase [33] and remarks the weak influence of the substrate that, for stronger molecule-substrate interactions, would tend to flatten the polymer structure. The observed replica in the ARPES spectrum may thus be related both to a twisted conformation of the polymer and to the commensuration with the substrate, since both effects are observed at the same time.

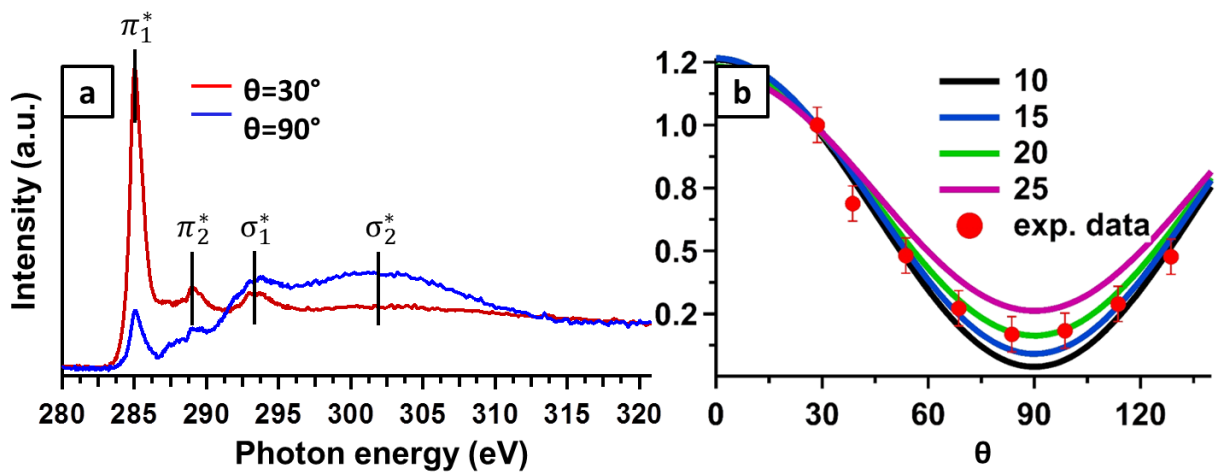

**Figure 4.** a) Angle-dependence of C K-edge NEXAFS spectra for the aligned PPP chains on Au(887). The spectra were collected with the polarization in the plane defined by the surface normal and the $[11\bar{2}]$ direction (perpendicular to the chains) and Θ is defined as the angle between the polarization and the surface normal (see Figure SI4). b) Angular dependence of the $\pi_1^*$ resonance intensity compared to the calculated behavior for different tilt angles of the phenyl rings. The calculated curves have been normalized to the first experimental value at θ=30° for better comparison.

Interestingly, the formation of this ordered adlayer leads to the observation of an intense replica of the gold *sp* band at 1.13 Å$^{-1}$, see the blue arrow in Figure 3d. Indeed, the adlayer introduces a new set of reciprocal lattice vectors $\vec{g}$, whose reciprocal length is 0.75 Å$^{-1}$ for the twisted PPP, which can lead to a back-folding of surface emission by a surface Umklapp process. Since the parallel component of the electron momentum is conserved in the photoemission process, inclusion of this new $\vec{g}$ gives rise, in the present case, to the observed band in the direction containing $\vec{k}_{\parallel} + \vec{g}$ (1.13+0.75 Å$^{-1}$ and folded back at ≈ 1.1 Å$^{-1}$, since the $\overline{K}$ point of gold is in the proximity of 1.49 Å$^{-1}$).[38,39] The presence of this Umklapp process, along with the disappearance of the surface state, hints to a significant hybridization between PPP and Au, which has been confirmed with DFT calculations (see Figure SI3). Nevertheless, the hybridization is still modest enough to allow for the observation of the full PPP band dispersion throughout the Au 3*d* band energies, down to -6.3 eV.

The effect of nitrogen incorporation into the conjugated PPP chain has been investigated by two pyridinic derivatives of 4,4"-dibromo-*p*-terphenyl, see monomers **2** and **3** in Figure 1. By means of these molecules, the changes in the electronic and transport properties can be mapped as a function of gradually increased nitrogen content. Again, by employing the 1D templating effect of the Au(887) vicinal surface, the targeted linear products have been obtained with a massive yield. STM imaging reveals well-oriented wires parallel to the steps direction and organized according to the same commensurate epitaxial matrix [4 2,0 3] of the undoped PPP chains. In fact, the pyridinic derivatives are characterized by the same distance and dihedral angle between neighboring rings (Fig. S6). Even if these PPP derivatives are isostructural with the undoped PPP, a closer inspection reveals an intramolecular modulation of the probed density of states with the same periodicity of the pyridinic functionalization as visible in the small scale STM images in Figure 5a and 5b.

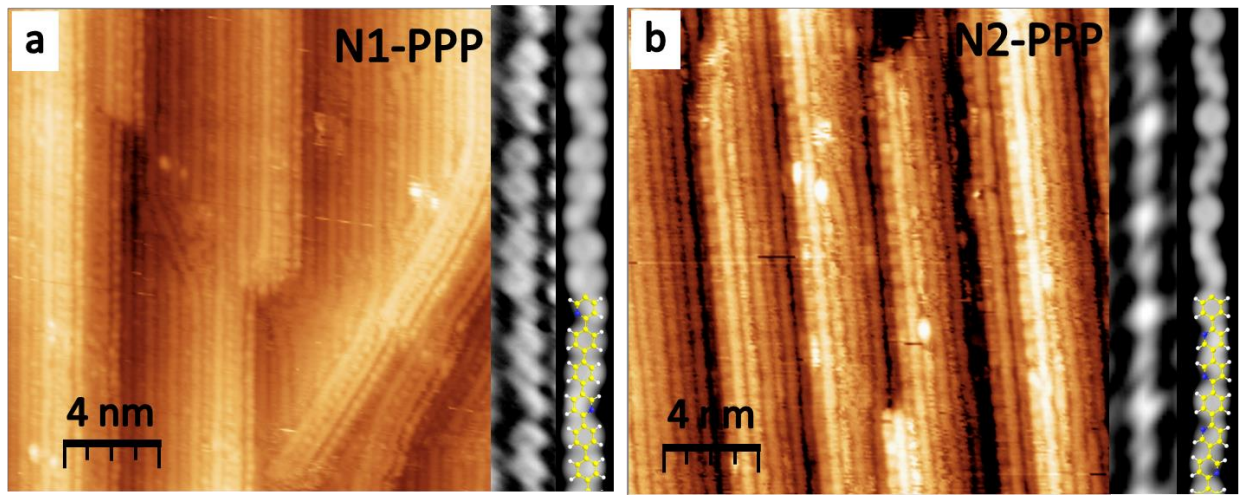

**Figure 5.** STM pictures of a) N1-PPP and b) N2-PPP grown on Au(887). The comparison between the experimental and DFT simulated STM images of a single are reported as lateral inset for the different polymers. a) 20x20 nm$^2$ V=-0.40 V I=0.51 nA; inset) V=-0.5 V I=2.0 nA; b) 20x20 nm$^2$ V=-0.71 V I=2.6 nA; inset) V=-0.5 V I=1.50 nA. Model color code: C yellow, N blue, H white.

DFT simulated STM images reported in Figure 5 correctly reproduce this behavior, where the pyridine rings appear smaller than phenyl rings, due to the lack of peripheral H atoms and to the lower N-related partial density of states near the Fermi energy (as reported below in Figure 6). Further evidence for the chemical environment of the nitrogen atoms comes from core-level photoemission (see SI for the complete XPS characterization of N 1$s$, C 1$s$, Br 3$d$ core levels). Moving to ARPES, the electronic structure probed on nitrogen-containing polymers share the main characteristics of PPP. In fact, the double periodicity and the Umklapp process are clearly visible both for N1-PPP and for N2-PPP. However, the VBMs exhibit a dependence on the degree of nitrogen substitution. The VBM for the pristine PPP is $E_F$-1.04 eV and decreases in the N1-PPP to $E_F$-1.26 eV and to $E_F$-1.38 eV in N2-PPP (see Figure 6a). Hence, we observe a continuous downshift of the VBM as the nitrogen content increases by as much as ≈ 0.35 eV for the doubly substituted PPP derivative. Since the conduction band remains above the Fermi level even in the doped wires, no information can be obtained about the magnitude of the band gap from these ARPES measurements, except that it should be larger than 1.4±0.1 eV.

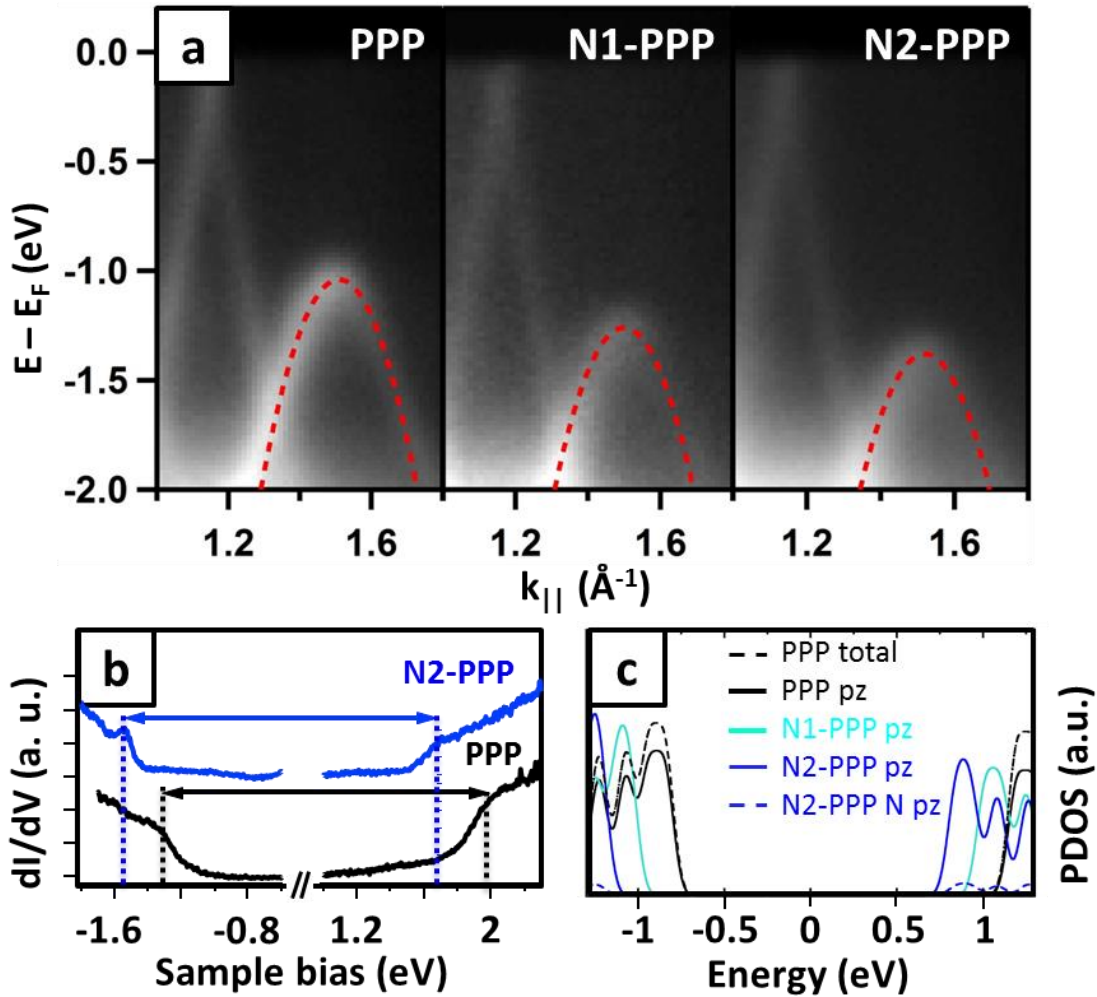

**Figure 6: Doping level-dependent electronic properties.** Panel a) reports the ARPES spectra of the different polymers at $k_\parallel$ close to the VBM. The superimposed red-dashed lines are the parabolic fit of the band revealing an effective mass of m*=0.19 $m_0$; b) STM dI/dV spectroscopy performed on PPP and N2-PPP, black and blue spectra respectively; c) computed projected densities of states (PDOS) of the different polymers. Total and $p_z$ projected density of states are compared for PPP, and Nitrogen-related $p_z$ is additionally disclosed for N2-PPP to underline the negligible contribution of N to valence and conduction bands.

Direct measurement of the band gaps has been performed by STS (dI/dV) along the polymer chains, see Figure 6b. In the experimental bias range of -1.8 V to +2.3 V, PPP clearly shows two band onsets around -1.3 V and +2.0 V, whereas the bands of the nitrogen-containing polymer N2-PPP rigidly shift toward lower bias voltages: about -1.6 V and +1.7 V. This implies a similar band gap (peak-to-peak) of $E_g$=3.3±0.2 eV for both PPP and N2-PPP. This value is in good agreement with the band gap reported for PPP on Cu(111) (3.1±0.2 eV),[19] albeit in stark contrast with that reported for PPP on Cu(110) (1.15 eV), where a massive band gap renormalization occurs due to a strong hybridization with the substrate.[40]

While STS and ARPES measurements reveal similar shifts upon N-doping, the onset of the valence band appears at about 0.25 eV higher binding energies in STS than in ARPES, probably related to the difference in the measured systems: STS was performed on single PPP chains on Au(111), while ARPES measurements were taken from compact PPP layers on Au(788). These differences in the systems may result in the observed shift by two possible mechanisms: i) a difference in work function. Indeed, we measure a 0.27 eV lower work function on Au(111) as compared to Au(788) (see Figure S12). Besides, since single chains are not flanked by Br atoms, changes in the local work function due to the absence of Br may add up to that difference.[41] And because a lower work function generally brings about higher binding energies (although not necessarily by the same amount[42]), this could explain our observations. ii) A lack of screening from neighboring PPP chains or Br atoms during STS characterization may also result in higher binding energies and thereby contribute to the observed differences.[43,44]

Most importantly, our STS characterization evidences that the downward shift of the valence band upon nitrogen incorporation is not related to a modified band gap, but to a shift of the whole band structure. This is further corroborated with the indirect evidence of a lower energy conduction band (CB) for N-doped polymers by the shifting onset of the NEXAFS resonances (Fig. S10), as well as with DFT calculations. The calculated projected densities of states (PDOS) of the different polymers are shown in Figure 6c. Apart from the offset of the conduction band and the absolute value of $E_g$, the DFT PDOS are qualitatively in agreement with our observations, since both the VB and the CB onsets shift rigidly towards lower bias voltage, leaving the band gap nearly unchanged. This effect is in agreement with the generally accepted picture of pyridine decreasing the electron density on the carbon ring by the electronegative nitrogen. Hence, the ionization potential and the electron affinity systematically increase, while the band gap is nearly unaffected, as observed for solid films of PPP (2.93 eV) and poly-pyridine (2.82-2.95 eV).[45,46] Within this picture it is also reasonable to assume that the shift is not strictly linear with N content, since the first N atom takes charge from a pure hydrocarbon molecule/polymer, while subsequent N atoms must take charge from an already charge-depleted carbon backbone of the molecules/polymers. This is indeed our observation, since going from PPP to N1-PPP, ARPES reveals a valence band downward shift of 0.22 eV, while adding one more N atom from N1-PPP to N2-PPP results in a lower shift of only 0.12 eV (0.34 eV with respect to the undoped PPP).[47]

Closely related studies on the effect of nitrogen doping have been performed on 6,11-dibromo-1,2,3,4-tetraphenyl-triphenylene precursors that afford chevron graphene nanoribbons after thermal activation.[15,48] One of these studies compared precursors with none, one and two incorporated N atoms, that is, with fractions of nitrogen heteroatoms within the doped precursors of 2.4% and 4.8%, respectively. The associated valence band shifts with respect to the undoped structure were 0.14 eV (2.4% heteroatoms) and 0.23 eV (4.8% heteroatoms). The other study compared precursors with none and four N atoms, revealing band shifts of 0.5 eV (9.5% heteroatoms). The valence band shift and fraction of nitrogen heteroatoms of the precursors used in this work correspond to 0.22 eV (5.6%, N1-PPP) and 0.34 eV (11.1%, N2-PPP), thus falling within a fully comparable range with respect to previous reports (Fig. S13).

However, as opposed to previous works, we additionally provide insight into transport properties, gathered in terms of the effective mass (m*) of holes in the valence band and derived from the curvature of the energy band. From a parabolic fit of the top of the valence band we determined m*=-0.19 $m_e$ for PPP, a value in agreement with DFT calculations on an infinitely long, unsupported PPP wire (-0.17 $m_e$) (see Figure S3) and in line with the high carrier mobility in PPP oligomers.[26] For direct comparison the same analysis was done on N1-PPP and N2-PPP, revealing a negligible effect of the pyridine ring, since both of them have m*=-0.19$m_e$. As indicated by the comparison of N2-PPP total $p_z$ and N-related $p_z$ projected DOS reported in figure 6c, the highest-lying valence bands are mainly composed by carbon π density, which amounts to saying that the heteroatom contributes little to define the hole carrier mobility in the valence band.[46,49]

**Conclusions**

In summary, we have grown three different extended one-dimensional polymers in a bottom-up fashion, namely poly-*para*phenylene and two pyridinic derivatives, with selected and gradually increasing nitrogen content. In order to characterize their band structure, macroscopically anisotropic samples have been prepared taking advantage of vicinal surface templating. Using angle resolved photoemission spectroscopy, we reveal a fully dispersive one dimensional behaviour of these semiconducting organic wires and that the electronic structure can be monotonically downshifted relative to the metal Fermi level as the pyridine substitution is increased within the molecular scaffold. Although the nitrogen insertion modulates the valence band onset by as much of 0.3 eV for the doubly doped monomer, DFT calculations in combination with scanning tunneling spectroscopy show that the band gap is unaffected ($E_g$= 3.3 ± 0.2 eV). Furthermore, from a parabolic fit of the ARPES data at the top of the valence band, a similar effective mass of $0.19m_0$ is evidenced for all polymers, implying comparable charge carrier mobility irrespective of the degree of nitrogen doping. This work thus confirms the possibility to tailor the electronic properties of functional organic nanowires, in particular the energy level alignment with respect to the Fermi level, without compromising the product's transport response (in terms of carrier mobilities) by selectively doping the molecular precursors. Finally, our findings represent a benchmark for the controlled growth of organic monolayers and the characterization of their electronic properties for future technological applications.

**Methods**

The Au(887) substrate surface was prepared by standard sputtering-annealing cycles. Subsequently, the samples were prepared by thermal evaporation of the precursors at 105°C, 100°C and 90°C for precursors **1**, **2** and **3**, respectively onto the substrate kept at room temperature. Multilayer deposition was followed by thermal annealing to 180 °C for 20 minutes, which brought about multilayer desorption and polymerization of the molecules at the interface with the metal. The samples thus obtained consisted of compact monolayers of polymers aligned along the substrate steps. STM and angle resolved ultraviolet photoemission spectroscopy (ARPES) were measured in a UHV system combining a commercial Omicron VT-STM and a homebuilt chamber equipped for ARPES with a closed-circuit He-compressor-cooled manipulator, a monochromatized gas discharge lamp and a SPECS Phoibos 150 electron analyzer. Being interconnected, STM and ARPES measurements were performed sequentially on the same sample without breaking the UHV conditions at any time. ARPES measurements were performed at a sample temperature of 90 K using the He I line (21.2 eV). The work function was determined from the low energy cut-off and Fermi edge in the photoemission spectra, measured on a sample polarized with 10 V. STS measurements were performed in a separated, commercial low-temperature STM operated at $LN_2$ temperature. Differently from the samples for ARPES, the polymer chains were grown at submonolayer coverage on a Au(111) surface. An electrochemically etched W tip was used for topographic and spectroscopic measurements. The spectroscopic dI/dV measurements were recorded by a lock-in amplifier modulating the sample bias with a sinusoidal voltage of 20 mV amplitude under open-feedback conditions. WSxM software was used to process all STM images.[50]

NEXAFS and XPS measurements were performed at the D1011 beamline at the MAX-IV laboratory (Lund, Sweden). The C K-edge NEXAFS spectra were measured in partial electron yield (PEY) mode. A retarding potential, $U_{ret}$, of −150 V was applied to the PEY detector in order to increase the signal-to-background ratio. The NEXAFS spectra were normalized to the spectra from the corresponding clean metal surfaces and to the continuum jump. The photoelectron spectra were measured relative to the Au $4f_{7/2}$ core level (BE = 84.00 eV) and were taken in a normal emission geometry. Fitting of the C 1s and Br 3d PE spectra was performed with the XPS peak software. The C 1s core level spectrum has been de-convoluted by assuming the presence of two different inequivalent carbon atoms within the molecular structure, i.e. carbon-bonded carbon, C-C, and nitrogen-bonded carbon, C-N and by keeping the ratio between the areas constant according to the theoretical stoichiometric ratios. The kinetic energy resolution of the SES-200 electron analyzer was set to 125 meV for the core-level spectra.

We simulated PPP N1-PPP and N2-PPP finite wires ~8nm long (18 rings), adsorbed on a Au(111) substrate, by means of Density Functional Theory in the mixed Gaussian plane waves framework as implemented in the CP2K code.[51] We used the Perdew−Burke−Ernzerhof (PBE) parametrization of the exchange− correlation functional.[52] Dispersion corrections were included following the recipe of Grimme and with a parameterization according to Ruiz.[53,54] The Au(111) substrate was modeled within the repeated slab geometry[55] using orthorhombic simulation cells containing four layers of gold and one layer of hydrogen atoms to suppress the surface state of Au(111) on one side of the slab. The surface of the slab corresponded to 240 unit cells (4 × 30 rectangular units, with extension ≈21 Å x 90 Å ), and more than 20 Å of vacuum was included. The atomic coordinates of the topmost two Au(111) layers and the adsorbed molecular species were fully optimized until forces on atoms were lower that $10^{-4}$ a.u. STM simulations were performed in the Tersoff−Hamann approximation[56] in constant-current mode using the toolkit asetk.[57] To overcome limitations of localized basis sets for non-periodic systems, the electronic states were extrapolated into the vacuum region.[58] Since the tip shape and the tip−sample distance were not accessible experimentally, a quantitative simulation of the experimental tunneling current was not attempted here. We chose the charge-density isovalue to provide a realistic tip−sample distance on the order of 5 Å. The PDOS presented in figure 6 are computed for wires in the gas phase constrained to the equilibrium adsorption geometry obtained from the calculations. A large simulation cell including the three ribbons together allowed us to avoid any problem of energy alignment.

ASSOCIATED CONTENT

**Supporting Information**. Epitaxy model and convoluted LEED pattern, second derivative ARPES data highlighting the full band dispersion, overlap of experimental and calculated band structure, additional STM images revealing typical chain lengths, representation of geometry in NEXAFS measurements, comparison of the angle dependent NEXAFS intensities on the three different polymers, C1s XPS data throughout polymer preparation procedure, C1s, Br3d and N1s XPS data comparing the various polymers studied, C K-edge NEXAFS spectra comparing the various polymers studied, summary of the energy level shifts between the various polymers studied as observed with different techniques, work function measurements from the low energy cutoff in photoemission spectra, graphical comparison of energy shifts upon N-doping reported for different systems, experimental details for chemical synthesis of precursors, and additional details regarding the calculations. This material is available free of charge *via* the Internet at http://pubs.acs.org.

AUTHOR INFORMATION

**Corresponding Author**

*d_g_oteyza@ehu.es  *francesco.sedona@unipd.it.

**Author Contributions**

The manuscript was written through contributions of all authors. All authors have given approval to the final version of the manuscript.

ACKNOWLEDGMENT

This work was partially funded by MIUR (PRIN 2010/11, Project 2010BNZ3F2: "DESCARTES") by EU project PAMS (agreement no. 610446), by the European Research Council (ERC) under the EU Horizon 2020 research and innovation programme (grant agreement No 635919), by the European Community's Seventh Framework Programme (FP7/2007-2013) CALIPSO under grant agreement nº 312284, by the Spanish Ministry of Science and Competitiveness (MINECO, MAT2013-46593-C6-6-P and MAT2013-46593-C6-4-P) and FEDER, by the Basque Government (Grant No. IT-621-13) and by the University of Padova (Grant CPDA154322, Project AMNES). We acknowledge the Swiss Supercomputing Center (CSCS) for computational support (project s507). Alberto Verdini is acknowledged for useful discussions on NEXAFS measurements and beamline staff at Max IV's D1011 for support during the beamtime.

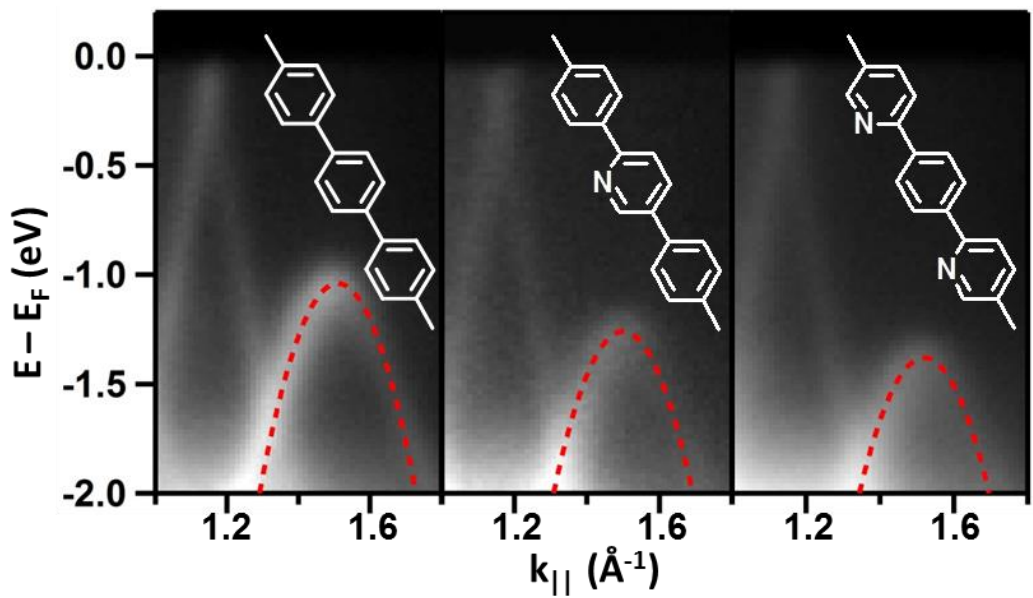

**Table of Contents.** Polymer doping level shifting the valence band maximum while preserving the effective mass and hence the charge transport response in terms of carrier mobility.